\documentstyle[preprint,revtex]{aps}
\newcommand{\bcn}{\begin{center}}
\newcommand{\beq}{\begin{equation}}
\newcommand{\beqn}{\begin{eqnarray}}
\newcommand{\ecn}{\end{center}}
\newcommand{\eeq}{\end{equation}}
\newcommand{\eeqn}{\end{eqnarray}}
\newcommand{\eq}[1]{Eq.~(\ref{#1})}
\begin{document}
\begin{small}
%
\begin{title}
 Gauge covariance and the fermion-photon vertex in \\
three- and four- dimensional, massless quantum electrodynamics
\end{title}
\author{Conrad J. Burden}
\begin{instit}
Department of Theoretical Physics, Research School of Physical Sciences and
Engineering,\\
Australian National University, GPO Box 4, Canberra, ACT 2601, Australia
\end{instit}
\author{Craig D. Roberts}
\begin{instit}
Physics Division, Argonne National Laboratory, \\
9700 South Cass Avenue, Argonne, Illinois 60439-4843
\end{instit}
\begin{abstract}
In the quenched approximation, the gauge covariance properties of three vertex
Ans\"{a}tze in the Schwinger-Dyson equation for the fermion self energy are
analysed in three- and four- dimensional quantum electrodynamics.  Based on the
Cornwall-Jackiw-Tomboulis effective action, it is inferred that the spectral
representation used for the vertex in the gauge technique cannot support
dynamical chiral symmetry breaking.  A criterion for establishing whether a
given Ansatz can confer gauge covariance upon the Schwinger-Dyson equation is
presented and the Curtis and Pennington Ansatz is shown to satisfy this
constraint.  We obtain an analytic solution of the Schwinger-Dyson equation for
quenched, massless three-dimensional quantum electrodynamics for arbitrary
values of the gauge parameter in the absence of dynamical chiral symmetry
breaking.\vspace*{30mm}

\hspace*{-2.5\parindent}{\it Preprint Number}~: PHY-7143-TH-93

\end{abstract}
%
\section{Introduction}
The Schwinger-Dyson Equations (SDEs) provide a valuable non-perturbative tool
for studying field theories. Phenomena such as confinement and dynamical chiral
symmetry breaking, which cannot be explained by perturbative treatments, can be
understood in terms of the behaviour of particle propagators obtained by
solving non-linear integral equations. However, the full set of SDEs for
any particular field theory contains an infinite tower of equations and
is thus intractible. A common approach for dealing with gauge field theories
is to approximate the fermion-gauge boson vertex by a suitable Ansatz
depending only on the dressed single particle propagators. The problem
is then reduced to that of solving a finite set of coupled equations for
the fermion and gauge boson propagators.

Ideally, of course, one would solve the SDE for the vertex itself.  However,
this equation involves the kernel of the fermion-antifermion Bethe-Salpeter
equation which cannot be expressed in a closed form; i.e., the skeleton
expansion of this kernel involves infinitely many terms.  Some approximation or
truncation of the system must therefore be made at a very early stage.  An
effective way to do this is to make an Ansatz for the vertex satisfying certain
criteria which the solution of the vertex equation must itself satisfy.  At the
present time this latter approach is the most efficacious manner in which to
proceed since it allows for a study of the relative importance of particular
vertex characteristics while avoiding the technical difficulties associated
with
solving the vertex equation directly.  However, we expect that it will soon be
necessary to study the vertex equation itself in order to make further
progress.

A primary purpose of this paper is to compare the effectiveness of three
commonly  used vertex Ans{\"a}tze, specifically with regard to their ability to
ensure the  gauge invariance of the theory. We begin with the requirement that
any  acceptable Ansatz, $\Gamma_\mu(p,q)$, must satisfy at least the following
criteria:
\begin{description}
\item[(a)] It must satisfy the Ward-Takahashi (WT) identity;
\item[(b)] It must be free of any kinematic singularities (i.e., expressing
        \mbox{$\Gamma_\mu(p,q)$} as a function of $p$ and $q$ and a functional
        of the fermion propagator, $S(p)$, then $\Gamma_\mu$ should have a
        unique limit as \mbox{$p^2\rightarrow q^2$});
\item[(c)] It must reduce to the bare vertex in the free field limit
            (i.e., when dressed propagators are replaced by bare propagators);
        and
\item[(d)] It must have the same transformation properties as the bare vertex,
        $\gamma_\mu$, under charge conjugation, $C$, and Lorentz
transformations
        (such as $P$ and $T$, for example).
\end{description}
Criterion (b) follows from Ref.~\cite{BC80} and criterion (c) is related to
this
since together they are necessary to ensure that the vertex Ansatz has the
correct perturbative limit.  The charge conjugation element of criterion (d) is
essential since it constrains the properties of \mbox{$\Gamma_\mu(p,q)$} under
\mbox{$p \leftrightarrow q$}.

One should also demand a further condition, namely that
\begin{description}
\item[(e)] Local gauge covariance should be respected.
\end{description}
In fact, a criticism of the SDE approach to solving gauge field theory has been
the apparent violation of gauge symmetry directly at the level of the equation
being addressed.  Ensuring gauge covariance of the solutions of the SDE goes
some way toward answering this criticism and allowing for a direct comparison
of
SDE results with those obtained from lattice gauge theory, for example.

Although condition (a) is a consequence of gauge invariance, it is only a
statement about the  longitudinal part of the vertex, and says nothing about
the
transverse part.  By itself it is insufficient to ensure contition (e)
\cite{BR91}. A well defined set of transformation laws which describe the
response of the propagators and vertex in quantum electrodynamics to an
arbitrary gauge transformation are given in an early paper by Landau and
Khalatnikov~\cite{LK56} (LK). These laws leave the SDEs and the WT identity
form-invariant and one can, in principle, ensure condition (e) by choosing an
Ansatz for $\Gamma$ which is covariant under the action of the LK
transformations.  Unfortunately, however, the transformation rule for the
vertex
is quite complicated, making this procedure difficult to implement.  Here we
will adopt a slightly different procedure. The LK transformation rule for the
fermion propagator is relatively straightforward, and we are able to check {\it
a posteriori} whether solutions for propagators obtained from a particular
vertex Ansatz transform appropriately.

Herein we discuss three- and four- dimensional, Euclidean, quenched quantum
electrodynamics (QED$_3$ and QED$_4$, respectively) and when discussing both we
choose to work with four-component spinors~\cite{Pi84}. (In formulating the
theory in Euclidean space we adopt the strategy of Ref.~\cite{Eqft}.)  In
describing the theory as ``quenched'' we mean that fermion loop contributions
to
the photon propagator are ignored; i.e., vacuum polarisation corrections are
neglected.

We remark that QED$_3$ has been much studied in recent years because of its
similarities with quantum chromodynamics (viz.  confinement and chiral symmetry
breaking), because its dimensioned coupling provides a natural scale which
makes
it a useful tool for modelling theories relevant to unification and because it
is not plagued by ultraviolet divergences.
For our purposes, however, it is the fact that in both QED$_3$ and QED$_4$ the
fermion SDE is solved by the combination of bare vertex and bare fermion
propagator that makes these theories interesting. The LK transform of the bare
fermion propagator from Landau to any other covariant gauge is readily found.
Any Ansatz for the vertex which does not admit the transformed propagator as a
solution  for an arbitrary value of the gauge parameter can  then be eliminated
as a possible candidate and is unlikely to form  a basis for a gauge covariant
vertex in  realistic models of non-Abelian theories.

We describe the vertex Ans\"{a}tze we are considering in detail in Sec.~II.
The
Ansatz of Ref.~\cite{H91} is equivalent to that employed in recent studies of
the SDE using the gauge technique~\cite{DW77,WD92}.  (The ``gauge technique''
assumes that the elements of the SDEs, propagators, etc., have spectral
representations in terms of which the SDEs are reformulated and then solved for
directly.)  We show in Sec.~III, using the Cornwall-Jackiw-Tomboulis effective
action~\cite{CJT74} (of which the fermion SDE can be interpreted as the
Euler-Lagrange stationary point equation), that this vertex Ansatz cannot
support dynamical chiral symmetry breaking simply because it leads to
independent equations for the vector, \mbox{$\sigma_V$}, and scalar,
\mbox{$\sigma_S$}, pieces of the fermion propagator, \mbox{$S(p)= -i\gamma\cdot
p\; \sigma_V(p) +\sigma_S(p)$}; the equation for $\sigma_S$ being homogeneous.
This is true of any Ansatz that yields independent equations for
\mbox{$\sigma_V$} and \mbox{$\sigma_S$} in the chiral limit.  (This is
exemplified in the QCD model of Refs.~\cite{BRW92}.)

In Sec.~III we discuss the fermion SDE in QED$_3$ and QED$_4$ in some detail
and
give numerical solutions to the QED$_3$ fermion SDE for various vertex
Ans{\"a}tze. In these studies we concentrate mainly on the case of no dynamical
mass generation (although the vacuum of massless QED$_3$ is generally believed
to be chirally asymmetric~\cite{Pi84,DKK90}, as may be that of quenched
QED$_4$~\cite{RC86}) and demonstrate analytically that the vertex Ansatz
proposed in Ref.~\cite{CP90} leads to a SDE which is solved by the LK transform
of the bare vertex. The remaining two Ans{\"a}tze, however, do not satisfy this
test. The observation of LK covariance enables us to obtain an analytic
solution
to the quenched, massless SDE in QED$_3$ for arbitrary values of the gauge
parameter in the absence of dynamical chiral symmetry breaking.  We summarise
our results and conclusions in Sec.~IV.  In an appendix we summarise the LK
transformations for QED and give the LK transformed three-dimensional free
massless fermion propagator for an arbitrary positive value of the covariant
gauge parameter.

\section{Fermion-photon vertices}

The most general form for a fermion-photon vertex satisfying criteria (a)
to (d) above has been given by Ball and Chiu~\cite{BC80} and, in Euclidean
space,
it can be written as follows:
\begin{equation}
\Gamma_\mu(p,q) = \Gamma_\mu^{\rm BC}(p,q) + \Gamma_\mu^{\rm T}(p,q),
                                  \label{gen}
\end{equation}
where
\begin{eqnarray}
\Gamma_\mu^{\rm BC}(p,q) & = & \frac{1}{2} \left[A(p)+A(q)\right]\gamma_\mu
\nonumber \\
& + & \frac{(p+q)_{\mu}}{p^2 -q^2}
\left\{ \left[ A(p^2)-A(q^2)\right]
                 \frac{\left[ \gamma\cdot p + \gamma\cdot q\right]}{2}
- i\left[ B(p^2) - B(q^2)\right]\right\} .
\label{BC}
\end{eqnarray}
and $\Gamma_\mu^{\rm T}$ is an otherwise unconstrained transverse
piece satisfying
\begin{equation}
(p-q)_\mu \Gamma_\mu^{\rm T}(p,q) = 0,
     \hspace{5 mm} \Gamma_\mu^{\rm T}(p,p) = 0.   \label{Tv}
\end{equation}
$\Gamma_\mu^{\rm BC}$ is given in terms of the dressed fermion propagator
\begin{equation}
S^{-1}(p) = i\gamma\cdot p A(p) + B(p),   \label{prop}
\end{equation}
where $A$ and $B$ are scalar functions of $p^2=p_\mu p_\mu$.
Our Euclidean space $\gamma$-matrices satisfy
$\{\gamma_\mu,\gamma_\nu\}= 2\delta_{\mu \nu}$.

Chiral symmetry breaking in QED$_3$, both in its quenched form \cite{BR91}
and in the presence of dynamical fermions~\cite{PW91},
has been studied with some success by
arbitrarily setting the tranverse part (\ref{Tv}) of the vertex equal to
zero. The remaining part, $\Gamma_\mu^{\rm BC}$, is the first of the three
Ans{\"a}tze we will consider herein. When dynamical mass generation
is also allowed for it goes some way towards ensuring that the chiral
condensate has only a weak dependence on the gauge parameter in
QED$_3$~\cite{BR91}, and provides
a value for the condensate in close agreement with that obtained from lattice
simulations~\cite{DKK90,B92}. However, our earlier numerical
studies~\cite{BPR92} have shown that the resultant fermion propagator
does retain some dependence on the choice of gauge.

The second vertex we consider is that proposed by Haeri~\cite{H91} which, in
Euclidean space, can be written as
\begin{eqnarray}
\Gamma_\mu^{\rm H}(p,q) & = &
i \left(\alpha_\mu S^{-1}(q) - S^{-1}(p)\alpha_\mu\right), \label{Hspec}
\end{eqnarray}
with \mbox{$\alpha_\mu =
[\gamma\cdot p \gamma_\mu + \gamma_\mu\gamma\cdot q]/[p^2-q^2]$}, or
alternatively:
\begin{equation}
\Gamma_{\mu}^{\rm H}(p,q) =
 \frac{p^2A(p) - q^2A(q)}{p^2-q^2} \gamma_\mu
         + \frac{A(p)-A(q)}{p^2-q^2} \not\!p \gamma_\mu \not\!q
- i \,
\frac{B(p)-B(q)}{p^2-q^2}(\gamma\cdot p \gamma_\mu + \gamma_\mu\gamma\cdot q).
                          \label{Ha}
\end{equation}
$\Gamma_{\mu}^{\rm H}$ is easily seen to satisfy criteria (a) to (d) and must
therefore be of the form Eq.~(\ref{gen}).

It is interesting to note that the Haeri vertex is identical
to the spectral representation of the vertex
employed in the gauge technique~\cite{DW77,WD92}:
\begin{equation}
S(p)\Gamma_\mu(p,q)S(q) = \int_{-\infty}^{\infty}d\omega \, \rho(\omega)
     \frac{1}{\not\!p - \omega} \gamma_\mu \frac{1}{\not\!q - \omega},
                           \label{spV}
\end{equation}
where $\rho(\omega)$ is the spectral density of the fermion propagator:
\begin{equation}
S(p) = \int_{-\infty}^{\infty}d\omega\, \frac{\rho(\omega) }
              {\not\!p - \omega}.  \label{spP}
\end{equation}
This result is true irrespective of whether the fermion aquires a mass and
can be easily verified by direct substitution and comparison (after
continuation
of \eq{Hspec} to Minkowski space)~\cite{M92}.

Thirdly we consider the Ansatz of Curtis and Pennington (CP).  In order to
ensure multiplicative renormalisability, they have proposed a vertex for which
the transverse part, $\Gamma_\mu^{\rm T_{CP}}$, takes the form~\cite{CP90,CP92}
\begin{eqnarray}
\Gamma_\mu^{\rm T_{\rm CP}}(p,q) & = &
  \frac{A(p)-A(q)}{2d(p,q)}
        \left[ \gamma_\mu (p^2-q^2)
       - (p+q)_\mu (\gamma\cdot p - \gamma\cdot q) \right],
\label{CPT}
\end{eqnarray}
with
\begin{equation}
d(p,q) = \frac{(p^2-q^2)^2 + [M^2(p) + M^2(q)]^2}{p^2 + q^2},
\end{equation}
where \mbox{$M = B/A$}, yielding the Ansatz
\begin{eqnarray}
\Gamma_\mu^{\rm CP}(p,q) & = &\Gamma_\mu^{\rm BC} + \Gamma_\mu^{\rm T_{\rm CP}}
\nonumber\\
      & = & \frac{p^2A(p) - q^2A(q)}{p^2-q^2} \gamma_\mu
       + \frac{(p+q)_\mu}{p^2-q^2} \frac{A(p)-A(q)}{p^2-q^2}
           (p^2 \gamma\cdot q - q^2 \gamma\cdot p) \nonumber \\
      & &   + \mbox{$B$-dependent parts}.
\label{CP}
\end{eqnarray}
For QED$_4$, the CP vertex gives a chirally symmetric fermion propagator
which is exactly multipicatively renormalisable at all momenta~\cite{CP91}.
It has also been used in Landau gauge QED$_3$ in conjunction with a one-loop
corrected photon propagator~\cite{CPW92}, with the result that chiral
symmetry is broken irrespective of the number of fermion flavours.


%
\section{The Quenched Schwinger-Dyson equation}
We now turn our attention to a consideration of the quenched fermion SDE for
QED$_3$ and QED$_4$:
\begin{equation}
1 = (i\gamma\cdot p + m)S(p) + e^2 \int \frac{d^{\rm d} q}{(2\pi)^{\rm d}}
      D_{\mu \nu}(p-q) \gamma_\mu S(q)\Gamma_\nu(q,p) S(p). \label{SDE}
\end{equation}
By quenched we mean that virtual fermion loops are ignored in the gauge boson
propagator which corresponds to setting \mbox{$\Pi(k)=0$} in Eq.~(\ref{CD}).
Our aim is to study the gauge covariance properties of Eq.~(\ref{SDE}) with the
Ans\"{a}tze for the vertices described above.  An Ansatz which leads to a
fermion propagator which does not respond to a gauge transformation in the
manner prescribed by the LK transformations, \eq{LKF}, can reasonably be
eliminated.  As we will see, this provides an additional constraint on the
transverse part of the vertex.

\subsection{Haeri Ansatz and Dynamical Chiral Symmetry Breaking}
We will first consider $\Gamma_{\mu}^{\rm H}$ of \eq{Ha}.  An interesting
observation is that, writing the propagator in the form
\begin{equation}
S(p) = -i\gamma\cdot p \sigma_V(p) + \sigma_S(p) \label{Svs}
\end{equation}
and defining the partially amputed vertex
\begin{equation}
\Lambda_{\mu}^{\rm H}(p,q) = S(p) \Gamma_{\mu}^{\rm H}(p,q) S(q),
\end{equation}
\eq{SDE} provides two decoupled equations, one for $\sigma_V$ and one for
$\sigma_S$, when $m=0$; i.e., for massless fermions.  This is obvious upon
inspection since $\Lambda_{\mu}^{\rm H}$ involves $\sigma_V$ multiplied only
with
odd numbers of $\gamma$ matrices and $\sigma_S$ multiplied only with even
numbers.  It is also worth noting that the equation for $\sigma_S$ is always
homogeneous and hence the solution is determined only up to an arbitrary
multiplicative constant.

With \eq{Ha} in \eq{SDE} one always has the chiral symmetry preserving solution
\begin{equation}
S^{\rm W}(p) = -i\gamma\cdot p \sigma_{V}^{\rm W}(p) \label{Csym}
\end{equation}
for $m=0$ and, in addition, it is also probable that the equation admits a
dynamical
chiral symmetry breaking solution for $m=0$ which would have the form
\begin{equation}
S^{\rm NG} = -i\gamma\cdot p \sigma_{V}^{\rm W}(p) + \sigma_{S}^{\rm NG}(p).
\label{Casym}
\end{equation}
This was the case, for example, in the phenomenological QCD studies of
Ref.~\cite{HH91}.  We remark that in \eq{Csym} and \eq{Casym} the vector part
of
the propagator is necessarily the same.  This is essential to the argument that
follows and is what sets this Ansatz apart for the others we consider.

The SDE is the stationary point equation for the CJT effective
action~\cite{CJT74} which, evaluated at this stationary point, is~\cite{KS85}:
\begin{equation}
V[S] = \int\frac{d^{\rm d}p}{(2\pi)^{\rm d}}
\left[{\rm tr}\ln[1-\Sigma(p) S(p)] + \frac{1}{2}{\rm
tr}[\Sigma(p)S(p)]\right].
\end{equation}
One might measure the relative stability of these extremals by evaluating the
difference \mbox{$V[S^{\rm NG}] - V[S^{\rm W}]$}.
For an Abelian gauge theory with $N_f$ flavours of fermion one finds (for $d=3$
or $4$ since we use $4$ component spinors) that
\begin{equation}
V[S^{\rm NG}] - V[S^{\rm W}]
 =  2 N_f \int\frac{d^{\rm d}p}{(2\pi)^{\rm d}}
\ln \left[ 1 + \frac{1}{p^2}
\frac{\sigma_{S}^{2}(p)}{\sigma_{V}^{2}(p)}\right]  >  0,
\end{equation}
since it is reasonable to assume that $\sigma_S$ and $\sigma_V$ are real for
real Euclidean  $p^2$.  (Since the equation for $\sigma_S$ is homogeneous, this
difference can, in fact, be made arbitrarily large:
\mbox{$\sigma_S\rightarrow \lambda\sigma_S$}.) Hence, based on the CJT
effective
action (which is the same as the auxiliary field effective action at the
stationary point) one finds that  $\Gamma_{\mu}^{\rm H}$ cannot support
dynamical chiral symmetry breaking.

\subsection{Chirally Symmetric Solution and Gauge Covariance}
For the remainder of this section we focus our attention on the chiral symmetry
preserving solution of the massless SDE:
\mbox{$ S(p) = -i\gamma\cdot p \sigma_V(p)$}.

We first note that since
\begin{equation}
\int d\Omega_{\rm d} \frac{1}{(p-q)^2}
\left( ({\rm d}-3)p\cdot q + 2\frac{p\cdot(p-q) (p-q)\cdot q}{(p-q)^2}\right)
\equiv 0 \label{angi}
\end{equation}
then, in Landau gauge, \eq{SDE} admits the free propagator solution
\begin{equation}
S(p) =  \frac{1}{i \gamma\cdot p}
\end{equation}
for each of the vertices discussed herein because of criterion (c).
We therefore immediately have the important result that if a given vertex
Ansatz
is to satisfy the gauge covariance criterion then, for arbitrary $\xi$, the
associated SDE must have the LK transform of the free field propagator as its
solution (\eq{LKF}).

In order to study this it is helpful to consider the massless SDE in
configuration space:
\begin{eqnarray}
\lefteqn{\delta^{\rm d}(x-y) = \gamma\cdot\partial^x S(x-y) +} \nonumber\\
 & & \!\!\!\!\!\!\!\!\!\!\!\!\!\!\!\!
e^2 \int d^{\rm d}z d^{\rm d}x' d^{\rm d}y'
\gamma_\mu \left( D_{\mu\nu}^{\rm
T}(x-z)+\partial^z\partial^z\Delta(x-z)\right)
 S(x-x') \Gamma_\nu(z;x',y')S(y'-y)
\end{eqnarray}
where we have explicitly divided the gauge-boson propagator into a sum of a
transverse, gauge independent piece, \mbox{$D_{\mu\nu}^{\rm T}$}, and
longitudinal, gauge dependent piece, $\Delta$.
Making use of the WT identity
\begin{equation}
\partial_\mu \Gamma_\mu(z;x',y') =
S^{-1}(z-y')\delta^{\rm d}(x'-z) - \delta^{\rm d}(z-y')S^{-1}(x'-z)
\end{equation}
and the identity
\mbox{$  \int_x \gamma\cdot\partial^x S(x,x')S^{-1}(x',z)  =
        \gamma\cdot\partial^x \delta^{\rm d}(x-z)~,$}
one obtains the massless SDE in the following form:
\begin{eqnarray}
\delta^{\rm d}(x-y) & = & \gamma\cdot\partial^x S(x-y) \nonumber \\
& - & e^2 \left\{
\int d^{\rm d}z [\gamma\cdot\partial^x \Delta(x-z)]\delta^{\rm d}(x-z)
        -[\gamma\cdot\partial^x \Delta(x-y)]\right\}S(x-y) \nonumber \\
& + & e^2 \int d^{\rm d}z d^{\rm d}x' d^{\rm d}y'
\gamma_\mu  D_{\mu\nu}^{\rm T}(x-z) S(x-x') \Gamma_\nu(z;x',y')S(y'-y).
\label{SDEx}
\end{eqnarray}
Now it is clear by inspection that if
\begin{equation}
\int d^{\rm d}z d^{\rm d}x' d^{\rm d}y'
\gamma_\mu  D_{\mu\nu}^{\rm T}(x-z) S(x-x') \Gamma_\nu(z;x',y')S(y'-y) = 0~;
\label{SDExc}
\end{equation}
then \eq{LKF}, with \mbox{$S(x;\xi=0)$} given in \eq{Fxfree}, is a solution of
the massless SDE; i.e., it is a solution if the last term on the right hand
side
of \eq{SDE} is identically zero in Landau gauge.

Most studies of the SDEs are undertaken in momentum space and it is a simple
matter to transcribe Eqs.~(\ref{SDEx}) and (\ref{SDExc}).  We see that the
solution of the SDE is LK covariant if
\begin{equation}
\int \frac{d^{\rm d}q}{(2\pi)^{\rm d}}\, D_{\mu \nu}^{\rm T}(p-q) \gamma_\mu
S(q) \Gamma_\nu(q,p)  =0,
\label{Ceprime}
\end{equation}
where $D_{\mu \nu}^{\rm T}(k) = (\delta_{\mu \nu} - k_\mu k_\nu/k^2)/k^2$ in
the
quenched theory, in which case the propagator satisfies:
\begin{equation}
1= i\gamma\cdot p S(p) + \xi e^2 \int \frac{d^{\rm d}q}{(2\pi)^{\rm d}}
\frac{i\gamma\cdot (p-q)}{(p-q)^4}\left[ S(p) - S(q)\right]
\label{SDEcov}
\end{equation}
in the covariant gauge fixing procedure.

It is now a simple matter to analyse the gauge covariance properties of our
vertex Ans{\"a}tze.

\subsubsection{Ball-Chiu Ansatz}
Using the BC vertex of \eq{BC}  the QED$_3$ SDE takes the form:
\begin{eqnarray}
A(p) - 1 & = & \frac{-e^2}{4\pi^2p^2} \int_0^\infty dq\, \frac{1}{A(q)}
    \left[ \xi \left(\frac{p^2A(p) - q^2A(q)}{p^2-q^2} -
  \frac{p^2A(p) + q^2A(q)}{2pq} \ln\left|\frac{p+q}{p-q}\right|\right)
               \right.         \nonumber \\
   & & \left.
  -\left(1 - \frac{p^2+q^2}{2pq}\ln\left|\frac{p+q}{p-q}\right|\right)
            (p^2+q^2)\frac{A(p)-A(q)}{p^2-q^2} \right],
                          \label{BCE}
\end{eqnarray}
while in QED$_4$ it is
\begin{eqnarray}
A(p) - 1 & = & \frac{e^2}{8\pi^2 p^2} \int_0^\infty dq\, \frac{q}{A(q)}
  \left\{ \xi  \left( A(p) \frac{p^2}{q^2} \theta(p-q)
                    + A(q) \frac{q^2}{p^2} \theta(q-p) \right) \right.
                          \nonumber \\
 & &  - \left. \frac{3}{4} \frac{A(p)-A(q)}{p^2-q^2} (p^2+q^2)
     \left( \frac{p^2}{q^2} \theta(p-q)
          + \frac{q^2}{p^2} \theta(q-p) \right) \right\},
\end{eqnarray}

It is clear that in neither of these equations is the right hand side
identically zero in Landau gauge ($\xi=0$) and therefore this vertex cannot
have
the correct LK transformation properties.  (This had already been established
numerically in Ref.~\cite{BR91} for QED$_3$.)

\subsubsection{Haeri Ansatz}
Using the Haeri Ansatz of \eq{Ha} we find the following form of the SDE in
QED$_3$:
\begin{eqnarray}
A(p) - 1 & = & \frac{-e^2}{4\pi^2p^2} \int_0^\infty dq\, \frac{1}{A(q)}
    \left[ \xi \left(\frac{p^2A(p) - q^2A(q)}{p^2-q^2} -
  \frac{p^2A(p) + q^2A(q)}{2pq} \ln\left|\frac{p+q}{p-q}\right|\right)
               \right.         \nonumber \\
   & & \left.
   + 2pq \ln\left|\frac{p+q}{p-q}\right| \frac{A(p)-A(q)}{p^2-q^2}
               \right],        \label{HE}
\end{eqnarray}
while in  QED$_4$ it takes the form:
\begin{eqnarray}
A(p) - 1 & = & \frac{e^2}{8\pi^2 p^2} \int_0^\infty dq\, \frac{q}{A(q)}
  \left\{ \xi  \left( A(p) \frac{p^2}{q^2} \theta(p-q)
                    + A(q) \frac{q^2}{p^2} \theta(q-p) \right) \right.
                          \nonumber \\
 & &  - \left. 3 \frac{A(p)-A(q)}{p^2-q^2}
     \left( p^2 \theta(p-q) + q^2 \theta(q-p) \right) \right\},
\end{eqnarray}
Again it is clear that the right hand side of these equations is not zero in
Landau gauge and hence this vertex cannot have the LK transformation
properties necessary to ensure gauge covariance.

\subsubsection{Curtis-Pennington Ansatz}
The CP vertex is a different matter.  The QED$_3$ SDE is
\begin{equation}
A(p) - 1 = \frac{-e^2\xi}{4\pi^2p^2} \int_0^\infty dq\, \frac{1}{A(q)}
           \left(\frac{p^2A(p) - q^2A(q)}{p^2-q^2} -
  \frac{p^2A(p) + q^2A(q)}{2pq} \ln\left|\frac{p+q}{p-q}\right|\right),
                          \label{CPE}
\end{equation}
in which the right hand side is clearly zero in Landau gauge.  Hence this
vertex, or at least that part of it which contributes to the SDE, has the form
necessary to ensure gauge covariance of the chirally symmetric fermion
propagator.

It is possible to solve this equation analytically.  The solution, for $\xi>0$,
is
\begin{equation}
\frac{1}{A(p)} =
  1 - \frac{e^2\xi}{8\pi p} \arctan \left(\frac{8\pi p}{e^2\xi}\right),
                  \label{Vee}
\end{equation}
as it should be since this corresponds to the LK transform of the massles free
fermion propagator in QED$_3$, as we show in the Appendix.  (To obtain this
result we first rewrote \eq{CPE} in the form
\begin{equation}
1 - \frac{1}{A(p)} = \frac{-e^2\xi}{8\pi^2p} \left\{ \int_0^\infty dq\,
\frac{q}{ A(q)}
     \frac{d}{dq}\left(\frac{1}{q}\ln\left|\frac{p+q}{p-q}\right|\right)
     -\frac{1}{p^2 A(p)}\int_0^\infty dq \,q
      \frac{d}{dq}\left(q\ln\left|\frac{p+q}{p-q}\right|\right) \right\}.
\end{equation}
Noting that the second integral in this equation is zero and using the identity
\begin{equation}
\frac{1}{\pi} \int_0^\infty dx \, \frac{x}{1+x^2}
      ln\left|\frac{a+x}{a-x}\right| = \arctan a,
\end{equation}
\eq{Vee} follows.)

The  SDE for QED$_4$ using the CP vertex is given in Ref.~\cite{CP91} and can
be
written formally as
\begin{equation}
A(p) - 1 = \frac{\xi \alpha_0}{4\pi p^2}
\int_{0}^{\infty} dq^2 \left[ \theta(p^2-q^2) \frac{q^2}{p^2}
+\theta(q^2-p^2)\frac{p^2}{q^2}\frac{A(p)}{A(q)} \right]\label{CPqed4}
\end{equation}
with $\alpha_0 = e^2/(4\pi)$.  In Ref.~\cite{CP91} this equation was
solved by introducing an upper bound on the $q^2$ integral.  The
actual form of the solution depends on the manner in which the divergent
momentum integral is regularised.  However, the fact that the right hand side
of
\eq{CPqed4} is proportional to $\xi$ does not.
This equation is, of course, \eq{SDEcov} for \mbox{d $=4$} and hence the CP
vertex also satisfies criterion (e) in QED$_4$.
\vspace*{\medskipamount}

To illustrate our discussion we present plots of numerical solutions for the
function $1/A(p)$ in QED$_3$ obtained from the BC and Haeri vertex equations,
(\ref{BCE}) and (\ref{HE}),  at $\xi=1$, Fig. 1, together with the CP vertex
solution, Eq.~(\ref{Vee}),  also at $\xi=1$. It is clear that the BC and Haeri
vertices do not give the correct LK transformed bare propagator as a solution
and so fail to maintain the gauge covariance of the SDE.

To close this section we remark that \eq{Ceprime} provides us with a much
needed
additional constraint upon the vertex function which, while not a full
implementation of criterion (e), nevertheless is a restriction on the form of
the transverse part of the vertex:
\begin{description}
\item[(e$^\prime$)] In the absence of dynamical chiral symmetry breaking; i.e.,
for $\sigma_S \equiv 0$, the vertex must be such that \eq{Ceprime} is
satisfied,
\end{description}
where \mbox{$D_{\mu\nu}^{\rm T}(k)$} is the transverse part of the quenched
photon propagator.
\section{Summary}
The Schwinger-Dyson equation (SDE) approach to the solution of a gauge field
theory provides an intuitively attractive manner in which to address this
problem and one which is less computationally intensive than lattice gauge
theory, for example.  A serious impediment to this application is the apparent
lack of gauge covariance in all SDE studies to the present.  In the fermion SDE
this can be traced to inadequacies in the structure of the
approximate/truncated
fermion--gauge-boson vertex used in these studies.   Addressing this violation
of gauge symmetry in QCD is made difficult by the presence of ghost fields,
however, progress can be made with Abelian theories.   In addition to being
interesting in their own right, the outcome of these studies can provide some
understanding of necessary characteristics that should be incorporated in the
construction of phenomenological, model SDEs for QCD.  The results we have
reported herein, which are summarised below, may be seen in this connection in
addition to standing alone as a contribution to understanding gauge covariance
in
QED$_3$ and QED$_4$.

We have considered three different Ans\"{a}tze for the vertex in the
quenched, massless QED$_3$ and QED$_4$ fermion SDE: 1) that due to Ball and
Chiu~\cite{BC80}; 2) that due to Haeri~\cite{H91}; and 3) that due to Curtis
and
Pennington~\cite{CP90}.

In considering Ansatz 2) we observed that it is identical to that employed in
the gauge technique and that, based on the CJT effective action, this Ansatz
cannot support dynamical chiral symmetry breaking since a solution with
no dynamically generated fermion mass; i.e., a solution of the form
\mbox{$S(p) = -i\gamma\cdot p \sigma_V(p)$}, is always dynamically favoured in
this case.  The feature of this vertex which entails this is the fact that it
yields decoupled equations for $\sigma_V$ and $\sigma_S$ (\eq{Svs}) when the
fermion bare mass is zero.  Whenever this is the case the CJT effective action
will predict that the chirally symmetric solution is dynamically favoured.

We obtained a necessary condition which must be satisfied by any vertex
Ansatz if it is to confer gauge covariance on the quenched QED$_3$ and QED$_4$
SDEs.
This condition is simple: the Ansatz must allow a free, massless propagator
solution in Landau gauge; i.e., \mbox{$S^{-1}(p) = i \gamma\cdot p$},
which provides a much needed constraint on the transverse piece of the vertex,
\eq{Ceprime}. Only if this is the case can the solution of the SDE respond to a
change in the gauge parameter as prescribed by the LK transformations; i.e, can
the solution be gauge covariant. Only  Ansatz 3) satisfies this constraint and
it
satisfies it both in QED$_3$ and QED$_4$.  In demonstrating this we obtained an
analytic solution of the quenched, massless QED$_3$ SDE for arbitrary values of
the gauge parameter in the absence of dynamical chiral symmetry breaking.

In going beyond the quenched approximation \eq{angi} is modified as follows:
\begin{equation}
\frac{1}{(p-q)^2} \rightarrow \frac{1}{(p-q)^2}\frac{1}{1+\Pi(p-q)},
\end{equation}
where $\Pi(p-q)$ is the photon polarisation scalar.  Subsequent to this
modification it follows that the free, massless particle propagator is not a
solution in Landau gauge for any vertex satisfying (c). In this case \eq{SDEx}
combined with gauge covariance, \eq{LKF}, does not require \eq{SDExc}.

We may thus conclude that Ansatz 3) has another desirable feature, in
addition to those dicussed in Refs.~\cite{CP90,CP92,CP91}: that of ensuring
gauge covariance of the quenched SDE, at least in the absence of dynamical mass
generation. The other two Ans\"{a}tze may be discarded since they manifestly
cannot allow gauge covariance in QED$_3$ or QED$_4$.  Hence, one may make the
inference that these two Ans\"{a}tze are less likely to provide a good starting
point in phenomenological SDE studies in QCD than Ansatz 3).
\acknowledgements
Some of the calculations described herein were carried out using a grant
of computer time and the resources of the National Energy Research
Supercomputer
Center. The work of CDR was supported by the Department of Energy, Nuclear
Physics Division, under contract number W-31-109-ENG-38.


\unletteredappendix{The Landau \& Khalatnikov transformations}

As pointed out in Section I, one would like to restrict
the vertex further by imposing condition (e), namely that the form of the
vertex Ansatz, stated in terms of the dressed propagators, should be
covariant under local gauge transformations. The gauge transformation laws
relating the propagators and vertex of QED to their Landau gauge
counterparts were first given by Landau and Khalatnikov~\cite{LK56}. These
rules
are most easily specified in coordinate space and we give below the
corresponding Euclidean space transformation laws.

In an arbitrary gauge, the photon propagator is modified from its
transverse, Landau gauge, form $D_{\mu \nu}(x;0)$ by the addition of
a longitudinal piece parameterised by an arbitrary function $\Delta$:
\begin{equation}
D_{\mu \nu}(x;\Delta) = D_{\mu \nu}(x;0)
        + \partial_\mu \partial_\nu\Delta(x).   \label{LKD}
\end{equation}
The corresponding rule for the fermion propagator is
\begin{equation}
S(x;\Delta) = S(x;0) {\rm e}^{e^2[\Delta(0) - \Delta(x)]}, \label{LKF}
\end{equation}
Where $e$ in the exponent is the gauge coupling constant. The rule
for the fermion-photon vertex is
\begin{eqnarray}
B_\mu(x,y,z;\Delta) & = &
    B_\mu(x,y,z;0) {\rm e}^{e^2[\Delta(0) - \Delta(x-y)]} \nonumber \\
    & + & S(x-y;0) {\rm e}^{e^2[\Delta(0) - \Delta(x-y)]}
        \frac{\partial}{\partial z_\mu} [\Delta(x-z) - \Delta(z-y)],
                                      \label{LKV}
\end{eqnarray}
where $B_\mu$ is the non-amputated vertex defined in momentum space in
terms of the amputated vertex $\Gamma_\mu$ by
\begin{equation}
B_\mu(p,q) = S(p)\Gamma_\nu(p,q)S(q)D_{\mu \nu}(p-q).
\end{equation}
One can check directly that these transformations leave the WT
identity and SDE form-invariant~\cite{BPR92}.

In the usual covariant gauge fixing procedure the photon propagator
takes the form
\begin{equation}
D_{\mu \nu}(k;\xi) = \frac{1}{k^2(1+\Pi(k^2))}
             \left(\delta_{\mu \nu} - \frac{k_\mu k_\nu}{k^2} \right)
        + \xi\frac{k_\mu k_\nu}{k^4},        \label{CD}
\end{equation}
which is obtained by taking $\Delta$ in Eq.~(\ref{LKD}) to be
\begin{equation}
\Delta(x) = -\xi \int \frac{d^{\rm d}k}{(2\pi)^{\rm d}} \frac{e^{-ik\cdot
x}}{k^4}.
\end{equation}
Within this set of gauges, one finds that in QED$_3$ the tranformation rule for
the fermion propagator, Eq.~(\ref{LKF}), becomes~\cite{BPRa}
\begin{equation}
S(x;\xi) = S(x;0) e^{-e^2\xi\left|x\right| / 8\pi}. \label{LKFX}
\end{equation}

The free massless propagator is \mbox{$S^{-1}(p;0) = i \gamma\cdot p$}
which corresponds to the following function in configuration space:
\begin{equation}
S(x;0) = \frac{\gamma\cdot x}{4\pi |x|^3}. \label{Fxfree}
\end{equation}
Applying \eq{LKF} one obtains the LK transformed function in an arbitrary
covariant gauge:
\begin{equation}
S(x;\xi) = \frac{\gamma\cdot x }{4\pi |x|^3}
              e^{-e^2\xi\left|x\right| / 8\pi}.
\end{equation}
For $\xi > 0$ one may evaluate the Fourier amplitude directly to obtain
\begin{equation}
S(p;\xi) = \frac{-i \gamma\cdot p}{p^2}
             \left[1 - \frac{e^2\xi}{8\pi p}
             \arctan \left(\frac{8\pi p}{e^2\xi}\right) \right].
                          \label{ATAN}
\end{equation}

For completeness  we give a formula for the LK transform
of the bare vertex $\gamma_\mu$ from Landau gauge to an arbitrary covariant
gauge. Using Eqs.~(\ref{LKD}), (\ref{LKF}) and (\ref{LKV}) one obtains the
transformation rule for the partially amputated vertex
\begin{equation}
\Lambda_\mu(p,q) = S(p)\Gamma_\mu(p,q)S(q).
\end{equation}
which is simply:
\begin{equation}
\Lambda_\mu(x,y,z;\Delta)  =
    \Lambda_\mu(x,y,z;0) e^{e^2[\Delta(0) - \Delta(x-y)]}. \label{LKL}
\end{equation}

If the vertex is equal to the bare vertex in Landau gauge,
\begin{equation}
\Lambda_\mu(p,q;0) = -\frac{\not\!p \gamma_\mu \not\!q}{p^2 q^2},
\end{equation}
one  finds that, for arbitrary $\xi$,
\begin{equation}
\Lambda_\mu(p,q;\xi) = \frac{-1}{16\pi^2}
      \gamma_\alpha \gamma_\mu \gamma_\beta
  \frac{\partial^2}{\partial p_\alpha \partial q_\beta}\int d^3x\,d^3y\,
       e^{i(p\cdot x - q\cdot y)} e^{-e^2\xi\left|x-y\right| / 8\pi}
         \frac{1}{|x|^3 |y|^3}. \label{LKBARE}
\end{equation}


%
\figure{This is a plot of \mbox{$1/A(p)$} as a function of $p$ in QED$_3$ with
$\xi=1$ (Feynman Gauge) and $e^2=1$. The solid line is \eq{Vee}, the analytic
solution expected from the LK transformation;  the numerical results are:
$\star
=$  Haeri Ansatz; $\triangle = $
Ball-Chiu Ansatz; and $\Diamond = $ Curtis-Pennington Ansatz.  Clearly, the
Curtis-Pennington Ansatz yields the correct solution.}
\end{small}

\begin{references}
\bibitem{BPRa} In ref.~\cite{BPR92} the sign in the exponent of the
               equivalent formula, Eq.(39),
               should be minus instead of plus.
\end{references}
\end{document}